# High-energy-density and superhard nitrogen-rich B-N compounds

Yinwei Li[1,2], Jian Hao[1], Hanyu Liu[2,3], Siyu Lu[2], John S. Tse[2,4]

[1] *Laboratory for Quantum Design of Functional Materials, School of Physics and Electronic Engineering, Jiangsu Normal University, Xuzhou 221116, China*

[2] *Department of Physics and Engineering Physics, University of Saskatchewan, Saskatoon, Canada, S7N 5E2*

[3] *Geophysical Laboratory, Carnegie Institution of Washington, Washington D.C. 20015, USA*

[4] *State Laboratory for Superhard Materials, Jilin University, Changchun, 130012, China*

The pressure-induced transformation of diatomic nitrogen into non-molecular polymeric phases may produce potentially useful high-energy-density materials. We combine first-principles calculations with structure searching to predict a new class of nitrogen-rich boron nitrides with a stoichiometry of $B_3N_5$ that are stable or metastable relative to solid $N_2$ and *h*-BN at ambient pressure. The most stable phase at ambient pressure has a layered structure (*h*-$B_3N_5$) containing hexagonal $B_3N_3$ layers sandwiched with intercalated freely rotating $N_2$ molecules. At 15 GPa, a three-dimensional $C222_1$ structure with single N–N bonds becomes the most stable. This pressure is much lower than that required for triple-to-single bond transformation in pure solid nitrogen (110 GPa). More importantly, $C222_1$-$B_3N_5$ is metastable, and can be recovered under ambient conditions. Its energy density of ~3.44 kJ/g makes it a potential high-energy-density material. In addition, stress–strain calculations estimate a Vicker's hardness of ~44 GPa. Structure searching reveals a new clathrate sodalite-like BN structure that is metastable under ambient conditions.

Triple-bonded diatomic nitrogen ($N_2$) has the highest bond energy of all diatomic molecules. Previous experiments have shown that molecular nitrogen can only be transformed into a singly bonded polymeric phase at pressures greater than 100 GPa [1-3]. The preservation of such a non-molecular structure and its recovery under ambient conditions would provide a useful high-energy-density material. However, polymeric N has never been formed at ambient pressure[1-3], precluding its practical applicability. Therefore, other nitrogen-rich compounds with substantially lower dissociation pressures of the triple N≡N bond have been sought such as $LiN_3$[4], $NaN_3$[5], a $CO-N_2$[6] mixture, and $C_3N_{12}$[7]. In particular, Raza *et al.* [6] reported a $CO–N_2$ system where the triple N≡N bond dissociated at 52 GPa.

Boron nitride (BN) is the only known stable compound in the BN system. It is isoelectronic to carbon, and forms analogues to various carbon structures (*h*-BN[8], *r*-BN[9], *c*-BN[10], *w*-BN[11], amorphous BN[12], BN nanotubes[13], and BN nanomesh[14]). These polymorphs exhibit exceptional mechanical, thermal, optical, and catalytic properties. They have featured in various efforts to find superhard materials, with polymorphs developed by substituting B and N for C in different carbon allotropes. Structural searches from first principles have also been used to develop superhard materials, uncovering a new family of thermodynamically meta-stable BN polymophs that includes *bct*-BN[15] (or *pct*-BN[16]), *Z*-BN[17], *P*-BN[18], *T*-BN[19], $cT_8$-BN[20], *I*-BN[21], *O*-BN[22, 23], *Pbca*-BN[24], $B_4N_4$[25], *M*-BN[26], *Z′*-BN[26] and $BC_8$-BN[26]. A common property shared between these BN polymorphs is their predicted superhardness, with estimated Vicker's hardnesses between 47 and 66.8 GPa.

*Ab initio* structural prediction can reliably search for unknown structures. A recent example is the prediction of $NaCl_3$ and $Na_3Cl$ compounds, which have been confirmed experimentally[27]. Experiments have also demonstrated that molecular $H_2$ can interact readily with closed-shell molecules at easily accessible pressures (<20 GPa) in the formation of van der Waals compounds, such as $SiH_4–H_2$[28] and $H_2S–H_2$[29]. If B-N compounds with unusual properties could be stabilized, they could be used in superhard and high-energy-density materials.

In this study, stable compounds in the $B_xN_y$ ($x$, $y$ = 1–6) system are investigated by *ab intio* structure prediction[30, 31]. In addition to the known compound BN, several thermodynamically stable and metastable BN and $B_3N_5$ compounds are found. The newly predicted $B_3N_5$ polymer is a hard and high-energy-density material that is metastable under ambient temperature and pressure, and thus might be recoverable.

Structure predictions for $B_xN_y$ are performed using the particle swarm optimization technique implemented in the CALYPSO code [30, 31] and the Vienna *ab initio* simulation package[32]. CALYPSO has been used to investigate a great variety of materials at high pressures [33-38]. Detailed information on the calculations is provided in the Supplemental Material[39].

We first perform structure searches on $B_xN_y$ ($x$, $y$ = 1–6) at ambient pressure. Figure 1 summarizes the formation enthalpies calculated at a high level of accuracy and normalized on a per-atom basis for the most energetically favorable of the structures. The stable compositions form a convex hull, where a point lying on the tie line corresponds to a thermodynamically stable phase. As expected, BN has the lowest formation enthalpy. On the B-rich side, all the stoichiometries have significant negative enthalpies with respect to dissociation into elemental B and N. However, the enthalpies are all above the tie line connecting BN and B. Therefore, these B-rich polymorphs are only thermodynamically metastable, and are susceptible to decomposition into *h*-BN and B. The N-rich region includes a thermodynamically stable compound with stoichiometry of $B_3N_5$; the enthalpy of the energetically most stable structure (*h*-$B_3N_5$) is nearly equal to 3(BN) + $N_2$ calculated with both the PAW-PBE and vdW-DF functionals.

Figure 2 shows 10 predicted low-enthalpy structures of BN at ambient pressure, which can be categorized as 2D layered, 3D, and cage-like structures. The most stable crystalline forms of BN predicted here have layered structures. These include the experimentally known *h*-BN [8] (with space groups *P*-6*m*2, *P*6$_3$/*mmc*, and $P\bar{3}m1$), *r*-BN[9] (space group $R\bar{3}m$), and two new structures with space groups $P\bar{6}m2$ (denoted as $P\bar{6}m2$-2 in Fig. 2) and $P3_1$. All six layered structures consist of flat planes of $B_3N_3$ hexagons (Fig. 2), with the only differences among them being the patterns in the stacking of the layers, which are *ABA*… in *h*-BN and *AAA*… in the new $P\bar{6}m2$-BN. The layers are held together by weak van der Waals forces with a large average interlayer distance of 4.2 Å based on the PAW-PBE calculation, which decreases to 3.5 Å when vdW-DF functionals are included. Consequently, the six layered structures possess nearly identical enthalpies.

Three meta-stable 3D structures—the experimentally observed *c*-BN, *w*-BN structures and the previously predicted *bct*-BN structure[15, 16]—are also found in the structure search (Fig. 2). We also find a sodalite structure with the cubic $Pm\bar{3}n$ space group at a higher negative formation enthalpy (Fig. 2). The $Pm\bar{3}n$ structure, which is constructed from planar $B_3N_3$ hexagonal faces with B–N bond lengths of 1.57 Å, is identical to that of *c*-BN. Eight hexagons in the unit cell are linked to each other by sharing B–N edges. Consequently, planar $B_3N_3$ hexagons and $B_2N_2$ squares form $B_{12}N_{12}$ sodalite-like cages (Fig. 2). The B and N atoms form a tetragonally bonded structure with both B–N–B and N–B–N forming bond angles of either 90° or 120°.

The $Pm\bar{3}n$ structure is stable at ambient pressure because all phonon frequencies are positive definite (Supplemental Material, Fig. S1[39]) and the calculated elastic constants[39] satisfy the Born stability criteria[40]. The results indicate that $Pm\bar{3}n$-BN is metastable and may be recoverable once formed. The electronic band structure calculated with the HSE06 functional shows that $Pm\bar{3}n$-BN is an insulator with a large indirect band gap of 5.9 eV (Supplemental Material, Fig. S3a[39]), which lies between those of *h*-BN (5.2 eV) and *c*-BN (6.4 eV). A sodalite-like $B_{12}N_{12}$ nanocage cluster has been synthesized[41], which is the building block of the $Pm\bar{3}n$ structure. Interestingly, a similar sodalite-like cage structure has been proposed in carbon[42], and is regarded as the best candidate structure for synthesized "superdense" carbon[43]. We found that at 0 GPa C-sodalite is 0.38 eV/atom higher in energy than diamond. In comparison, $C_{60}$ is 0.30 eV/atom higher in energy than graphite. In this case, the total energy difference of $Pm\bar{3}n$-BN with respect to $c$-BN is only 0.26 eV/atom. Therefore, if C-sodalite is indeed the synthesized "superdense" carbon, we believe that BN-sodalite can also be synthesized.

The stability of N-rich $B_3N_5$ is surprising. Its most stable form at ambient pressure has a layered structure (Fig. 3). Interestingly, the two energetically comparable $P\bar{6}2m$ and $Pm$ structures (referred to as $h$-$B_3N_5$ hereafter) are constructed from the same planar hexagonal BN layers sandwiched with $N_2$ molecules (($BN)_3N_2$). Each BN layer in $h$-$B_3N_5$ consists of a network of hexagonal $B_3N_3$ with B–N bond lengths of 1.45 Å, identical to those in $h$-BN. The N–N bond length of the $N_2$ molecule in the two layered structures is 1.11 Å, which is only slightly longer than that (1.07 Å[44]) of pure $N_2$ and much shorter than that (1.21 Å) of a typical N=N double bond in $N_2F_2$. Therefore, the intercalated $N_2$ in $h$-$B_3N_5$ is nominally triple bonded (N≡N).

Both the $P\bar{6}2m$ and $Pm$ structures have the same $AA...$ stacking of BN layers with $N_2$ molecules sandwiched between pairs of adjacent planes. The only difference between them is the orientation of the $N_2$ molecules, which are perpendicular to the layers in $P\bar{6}2m$ but parallel in $Pm$ (dashed rectangles in Fig. 3). Removal of $N_2$ molecules from the two structures results in a simple layered structure of $P\bar{6}m2$-BN (Fig. 2), suggesting that the $N_2$ molecules are loosely bound between the BN hexagonal layers. To test this suggestion, total energy calculations are performed by rotating the $N_2$ incrementally in the (110) and (1-10) planes (Fig. 3) while fixing each molecule's center of mass. Plots of the calculated energy versus $N_2$ orientation are compared in Fig. 4(a). The maximum energy change during the $N_2$ rotation is only 1.7 meV/atom, even smaller than the estimated energy barrier (6 meV/atom) of $H_2$ rotation in orientationally disordered $hcp$-$H_2$[45]. Therefore, the $N_2$ molecules in $h$-$B_3N_5$ are almost certainly freely rotating at finite temperature. Surprisingly, we find that the band gap of $h$-$B_3N_5$ changes appreciably with the $N_2$ orientation: from 4.74 eV in the $P\bar{6}2m$ structure to 5.02 eV in the $Pm$ structure (Supplemental Material, Figs. S3b and S3c [39]).

In addition to $h$-$B_3N_5$, we find another layered structure with a $R\bar{3}m$ space group. In contrast to the free $N_2$ in $h$-$B_3N_5$, the $N_2$ units here participate in forming the puckered $B_3N_5$ layers containing edge-sharing $B_3N_3$ and $B_2N_4$ hexagons (Fig. 3). Consequently, the N–N bond distance of 1.48 Å is much longer than that in $h$-$B_3N_5$, and is closer in length to a single N–N bond. The B–N bond distance of 1.44 Å is similar to that in $h$-$B_3N_5$. The $B_2N_4$ hexagon is less stable than the $B_3N_3$ hexagon because the $R\bar{3}m$ structure has a much higher enthalpy than $h$-$B_3N_5$ (Fig. 3).

The structure search also yields an orthorhombic 3D $B_3N_5$ structure with the space group $C222_1$. This structure is metastable relative to $h$-$B_3N_5$ at 0 GPa, but is the most stable structure in the structure searches performed at 50 GPa. The calculated equation of states predicts that the phase transition from $h$-$B_3N_5$ to $C222_1$ occurs at 15 GPa (Fig. 4b). In the $C222_1$ structure, each B atom is tetragonally bonded to two different N atoms (N1 and N2) with B–N1 and B–N2 bond lengths of 1.58 and 1.53 Å, respectively. The $sp^3$ N1 is bonded to four B atoms, and $sp^2$ N2 is bonded to two B atoms and one N2 atom; both combine to form planar $B_2N_4$ hexagons with two equal N–N bonds 1.33 Å in length. In the $C222_1$ structure, the N2–N2 bond of two $sp^2$ N atoms indicates a single bond. In contrast to the large band gap of $h$-$B_3N_5$ and the various crystalline forms of BN, the presence of $sp^2$ and $sp^3$ N atoms make $C222_1$-$B_3N_5$ a semiconductor with a small band gap of 0.775 eV (Supplemental Material, Fig. S3d[39]).

The calculated phonon dispersions (Supplemental Material, Fig. S1b [39]) and elastic constants (Supplemental Material, Table S3[39]) confirm that $C222_1$-$B_3N_5$ is dynamically and mechanically stable at ambient pressure. Owing to the large difference in the bond strengths between single N–N (160 kJ/mol) and triple N≡N (954 kJ/mol) bonds[2], the transformation from polymeric nitrogen back to diatomic nitrogen is expected to be highly exothermic. The energy involved in the decomposition of $C222_1$-$B_3N_5$ into solid 3BN and gaseous $N_2$ at ambient pressure was calculated. The contribution of $PV$ to the enthalpy of gaseous $N_2$ was considered in view of its large volume (22.4 L/mol) under ambient conditions. At the PBE-GGA level, the energy difference of 3.57 eV corresponds to an energy density of ~3.44 kJ/g, which is lower than that predicted for *cg*-N (9.7 kJ/g), but higher than that predicted recently for CO–$N_2$ (2.2 kJ/g) [6]. Remarkably, $C222_1$-$B_3N_5$ is thermodynamically stable at 15 GPa, which is much lower than the formation pressures of *cg*-N (110 GPa) and polymeric CO–$N_2$ (52 GPa). This stability at a relatively low pressure suggests that $C222_1$-$B_3N_5$ might be synthesized by high-pressure high-temperature techniques. If it can be made, the compound will be a good high-energy-density material. The energy and phonon calculations predict two BN polymorphs (sodalite-like $Pm\bar{3}n$-BN and $C222_1$-$B_3N_5$) that may be quenched and recovered at ambient pressure. The stabilities of both compounds at ambient pressure and temperature are investigated by metadynamics simulations at 1 bar and 300 K (Supplemental Material, Fig. S2[39]). No structural changes are observed after 200 metasteps, indicating that both compounds are metastable under ambient conditions.

A strong three-dimensional covalent bond network is a key feature of superhard materials. The hardness values of $Pm\bar{3}n$-BN and $C222_1$-$B_3N_5$ were estimated with the microscopic hardness model[46-48], which calculates the hardness of a covalent crystal as $H_v = 740 P^u (v_b^u)^{-5/3}$, where $P^u$ and $v_b^u$ are the Mulliken overlap population and the volume of *u*-type bonding, respectively. This model predicts remarkably high hardnesses for $Pm\bar{3}n$-BN (58.4 GPa) and $C222_1$-$B_3N_5$ (78.5 GPa) (Supplemental Material, Table S4). Their high hardnesses compared with *c*-BN (65 GPa) allow both to be classified as superhard materials. These encouraging results are verified through more exact *ab intio* stress–strain calculations for $C222_1$-$B_3N_5$. The results presented in Fig. 5 (a) show weakness in the <011> direction, with a ideal tensile strength of 56 GPa; therefore, the (011) planes allow easy cleavage. We then evaluate the shear stress response in the (011) planes and a ideal shear strength of 44 GPa is obtained in the (011)[100] shear direction. This suggests a theoretical hardness of 44 GPa for $C222_1$-$B_3N_5$, much lower than that estimated by the microscopic hardness model. The discrepancy is not surprising as this empirical model is known to exaggerate hardness for open-framework structures. Despite the hardness being lower than that of *c*-BN, quenched recovered $C222_1$-$B_3N_5$ may still be an industrially useful material.

In conclusion, we performed a systematic search for stable compounds in the BN system. In addition to BN, the search found a new stable N-rich compound with stoichiometry of $B_3N_5$, which at ambient pressure has a layered structure with freely rotating $N_2$ molecules intercalated between the layers. Layered $B_3N_5$ should transform into $C222_1$-$B_3N_5$ above 15 GPa with the dissociation of triple N≡N bonds into single N–N bonds. Therefore, the $C222_1$ phase is a potential high-energy-density material. Calculations also revealed $C222_1$-$B_3N_5$ to be superhard. These results provide a promising new area of synthesis for nitrogen-rich high-energy-density materials *via* the intercalation of $N_2$ molecules into the BN layers.

**Acknowledgements**

Y. L. and J. H. acknowledge funding support from the National Natural Science Foundation of China under Grant Nos. 11204111 and 11404148, the Natural Science Foundation of Jiangsu province under Grant No. BK20130223, and a project funded by the Priority Academic Program Development of Jiangsu Higher Education Institutions (PAPD). J.S.T. and H. L. thank the National Natural Science Foundation of China under Grant Nos. 11474126. All the calculations were performed using the Westgrid facility with an allocation to J.S.T. and the Plato machine at the University of Saskatchewan. Work at Carnegie was partially supported by EFree, an Energy Frontier Research Center funded by the DOE, Office of Science, Basic Energy Sciences under Award No. DE-SC-0001057 (salary support for H.L.). The infrastructure and facilities used at Carnegie were supported by NNSA Grant No. DE-NA-00006, CDAC. Structures were plotted with VESTA[49] software.

**Figure Captions**

**Fig. 1.** Convex hull diagram for the BN system at ambient pressure. Formation enthalpy, $\Delta H$, is defined as $\Delta H = H(B_xN_y) - (xH(B) + yH(N))$. Alpha-B phase and alpha-$N_2$ phase are used in the calculation. Only stoichiometries with formation enthalpies close to the convex hull are presented.

**Fig. 2.** Ten predicted low-enthalpy structures for BN at ambient pressure. Horizontal bars beside each structure represent formation enthalpies with respect to elemental B and N; the color represents the type of structure. Dashed vertical lines in the first six layered structures are shown to distinguish the stacking sequence of the layers.

**Fig. 3.** Illustrations of four predicted low-enthalpy structures of $B_3N_5$ at ambient pressure. Dashed rectangles in the left two structures represent the different orientations of $N_2$ molecules.

**Fig. 4.** (a) Total energies as a function of $N_2$ orientation at ambient pressure in (110) and (1-10) planes of $h$-$B_3N_5$. Insets in (a) illustrate $N_2$ rotations. Blue atoms represent rotated $N_2$ molecules. (b) Enthalpy curves of $C222_1$ and R3m structures relative to $h$-$B_3N_5$ as a function of pressure.

**Fig. 5.** (a) Calculated tensile stress–strain relations for $C222_1$-$B_3N_5$ in various directions. (b) Calculated shear stress–strain relations for $C222_1$-$B_3N_5$ in the (011) easy cleavage plane.

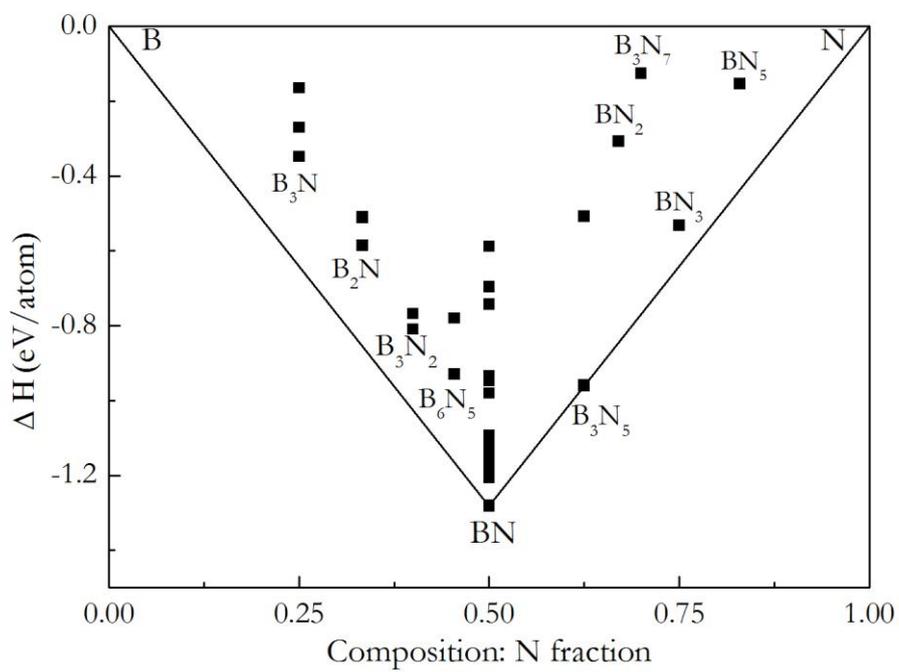

**Figure 1**

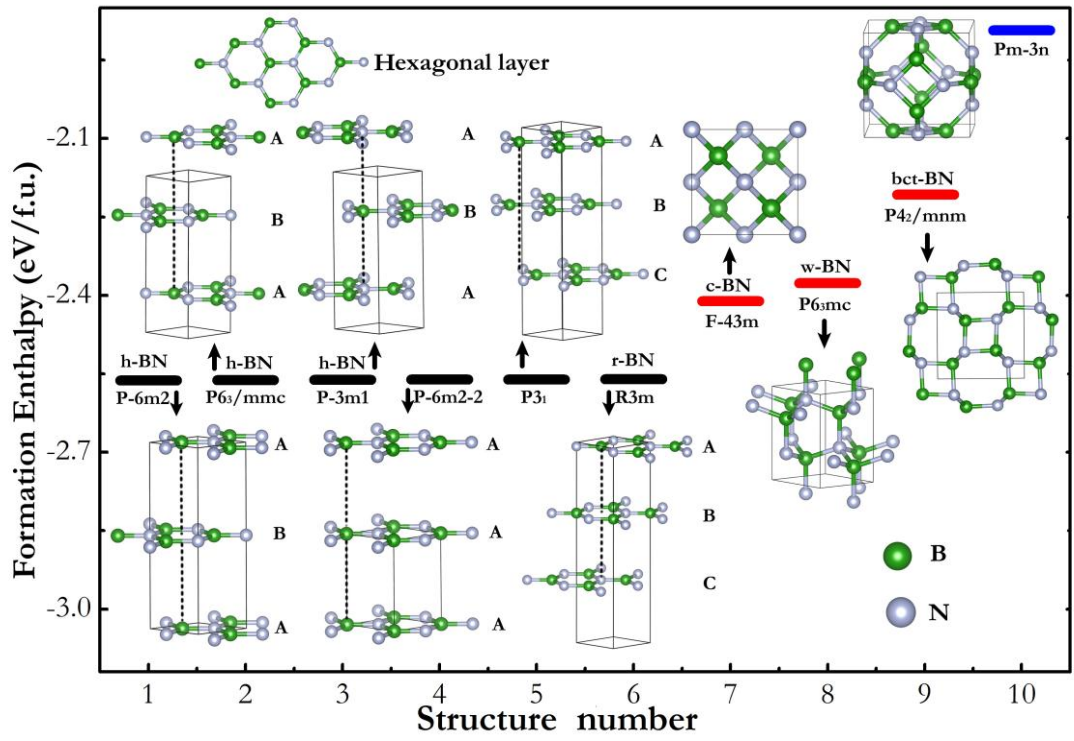

Figure 2

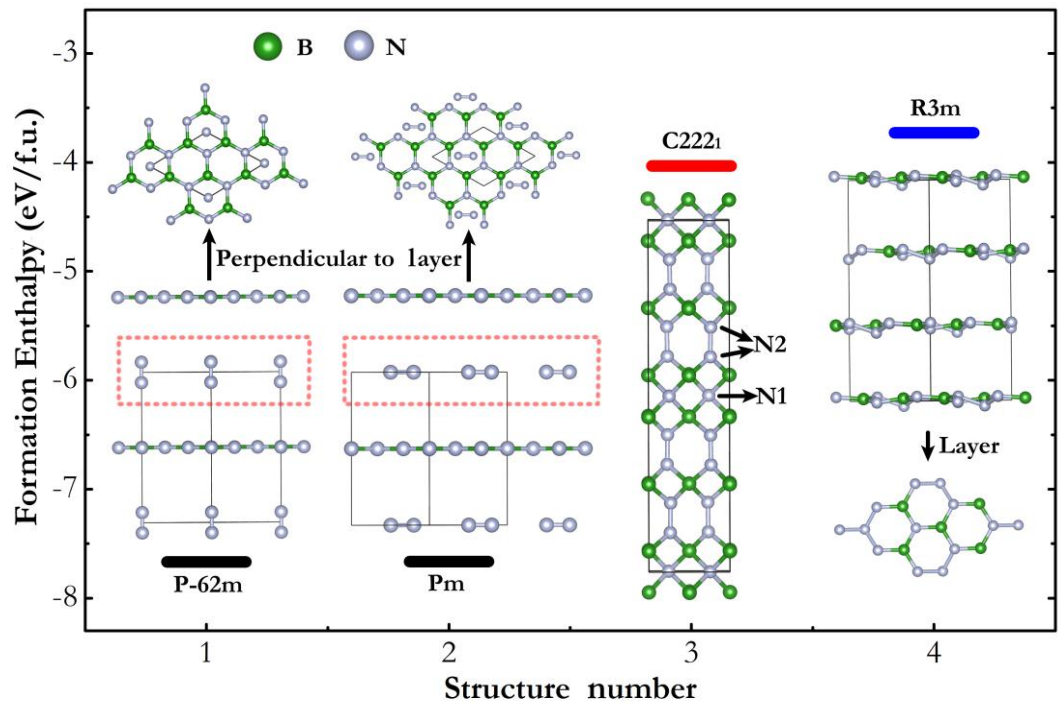

**Figure 3**

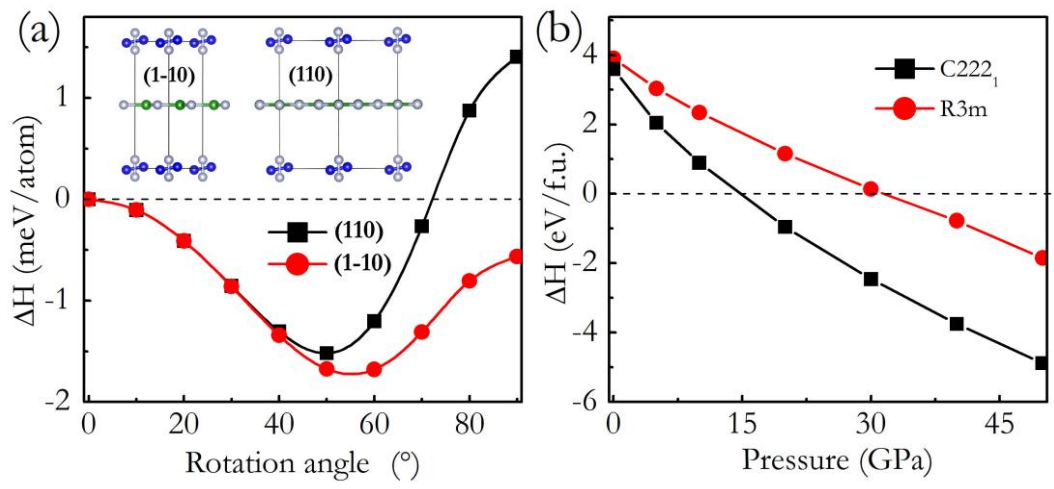

**Figure 4**

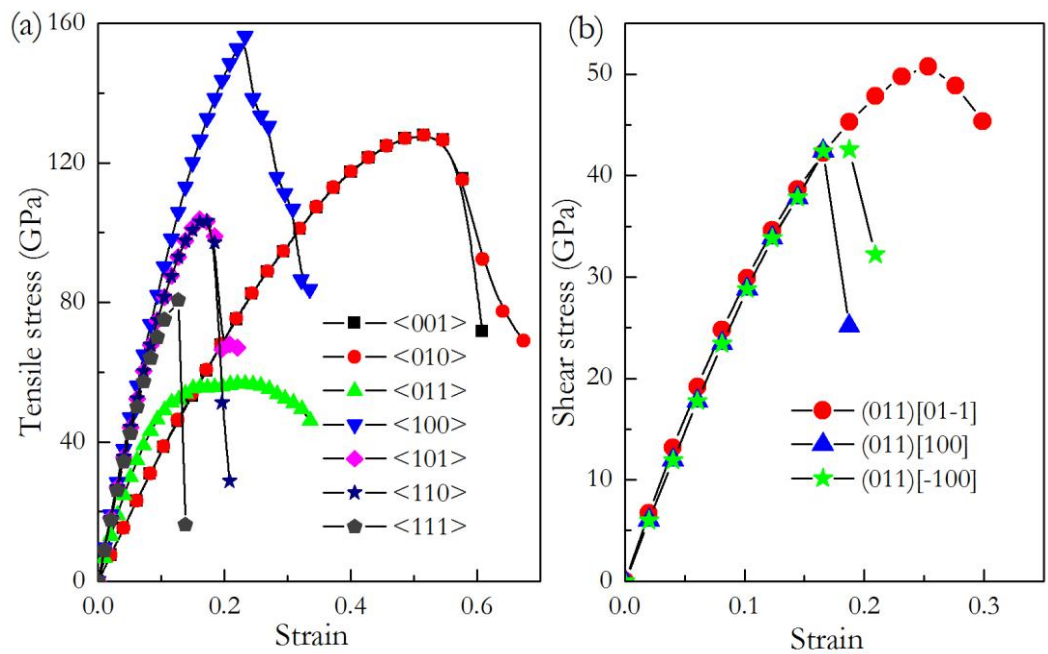

**Figure 5**